\documentclass[aps,prb,twocolumn,english,showpacs,superscriptaddress]{revtex4-1}
\usepackage[T1]{fontenc}
\usepackage[latin9]{inputenc}
\usepackage{rotating}
\usepackage{amsmath}
\usepackage{graphicx}
\usepackage{amssymb}
\usepackage{multirow}

\makeatletter

\DeclareFontEncoding{LGR}{}{}

\providecommand{\tabularnewline}{\\}

\@ifundefined{textcolor}{}
{%
 \definecolor{BLACK}{gray}{0}
 \definecolor{WHITE}{gray}{1}
 \definecolor{RED}{rgb}{1,0,0}
 \definecolor{GREEN}{rgb}{0,1,0}
 \definecolor{BLUE}{rgb}{0,0,1}
 \definecolor{CYAN}{cmyk}{1,0,0,0}
 \definecolor{MAGENTA}{cmyk}{0,1,0,0}
 \definecolor{YELLOW}{cmyk}{0,0,1,0}
 }

\usepackage{epsf}
\usepackage{epsfig}

\makeatother

\usepackage{babel}

\begin{document}

\title{Functionalization of BN Honeycomb structure by Adsorption and Substitution of Foreign atoms}
\author{C. Ataca}
\affiliation{Department of Physics, Bilkent University, Ankara 06800, Turkey} \affiliation{UNAM-Institute of Materials Science and
Nanotechnology, Bilkent University, Ankara 06800, Turkey}
\author{S. Ciraci} \email{ciraci@fen.bilkent.edu.tr}
\affiliation{Department of Physics, Bilkent University, Ankara 06800, Turkey} \affiliation{UNAM-Institute of Materials Science and Nanotechnology, Bilkent University, Ankara 06800, Turkey}

\begin{abstract} We carried out first-principles calculations within Density Functional Theory to investigate the structural, electronic and magnetic properties of boron-nitride (BN) honeycomb structure functionalized by adatom adsorption, as well as by the substitution of foreign atoms for B and N atoms. For periodic high density coverage, most of $3d$ transition metal atoms and some of group 3A, 4A, and 6A elements are adsorbed with significant binding energy and modify the electronic structure of bare BN monolayer. While bare BN monolayer is nonmagnetic, wide band gap semiconductor, at high coverage of specific adatoms it can achieve magnetic metallic, even half-metallic ground states. At low coverage, the bands associated with adsorbed atoms are flat and the band structure of parent BN is not affected significantly. Therefore, adatoms and substitution of foreign atoms at low coverage are taken to be the representative of impurity atoms yielding localized states in the band gap and resonance states in the band continua. Notably, the substitution of C for B and N yield donor and acceptor like magnetic states in the band gap. Localized impurity states occurring in the gap give rise to interesting properties for electronic and optical application of the single layer BN honeycomb structure.

\end{abstract}


\maketitle

\section{Introduction}

Research on BN based materials have grown gradually in recent years.\cite{McMillan, Pan, Miyoshi, Chopra, Arenal} This is not only due to their fascinating properties, such as hardness, high melting point, and large band gap, but also due to the geometric similarity of planar, two dimensional (2D) BN to graphene. Scientists already achieved the synthesis of single layer BN honeycomb structure on substrates\cite{Auwarter, Morscher} and a few layer thick structures from 3D hexagonal (h-)BN either on a substrate or freestanding.\cite{Han, Pacile, Novoselov} Recently, Jin \emph{et al.}\cite{Jin} reported the fabrication of freestanding BN honeycomb structure (we specify it simply as 2D BN throughout the text). The realization of the synthesis of 2D BN is rapidly attracting interest on BN, since it has 2D hexagonal lattice, which is commensurate to the lattice structure of covalently bonded graphene. More recently, the synthesis of single layer composite structures consisting of adjacent 2D BN and graphene domains is realized.\cite{natmat-bn-c} However, unlike semi-metallic graphene, 2D BN is a nonmagnetic, wide band gap semiconductor with an indirect energy gap of $4.64$ eV\cite{Topsakal} calculated within generalized gradient approximation (GGA). The indirect gap is further corrected to $6.82$ eV with GW$_0$ self-energy method by \c{S}ahin \emph{et al.}\cite{Sahin} A theoretical comparative study of 3D and 2D BN, and its nanoribbons comprising their mechanical, electronic and magnetic properties was reported by Topsakal \emph{et al.}\cite{Topsakal} 2D BN, and their nanoribbons can be easily functionalized by many different ways for different purposes such as doping\cite{Nose, hakem1, hakem2}, exchange of atoms and vacancies.\cite{Zobelli2, Orellana}

In this paper, using state-of-the-art first principles plane wave calculations we investigate the effects of adatoms adsorbed on 2D BN, as well as the substitution of foreign atoms for B and N atoms in the honeycomb structure. We consider both high coverage (where the coupling between adjacent foreign atoms is substantial) and low coverage (where the coupling is negligible). We conclude that at high coverage (or decoration) of specific adatoms one can turn the nonmagnetic, wide band gap material into magnetic, metallic or even half-metallic states. At low coverage, adatoms give rise to localized states in the band gap.

\begin{center}
\begin{figure}
\includegraphics[width=7.0 cm]{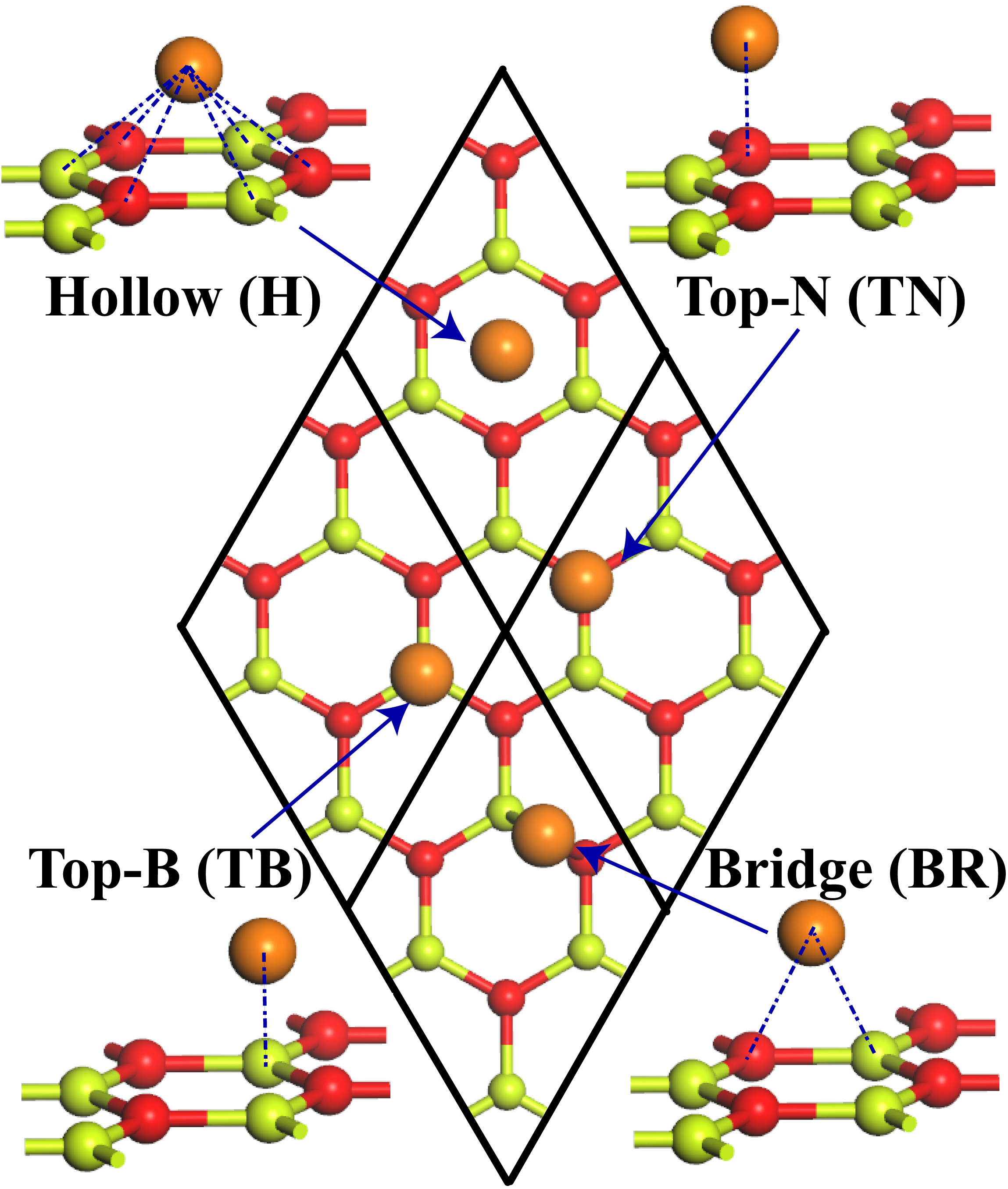}
\caption{(Color online) A (4x4) supercell of 2D BN honeycomb structure, which consists of four (2x2) supercells. Various possible adsorption sites of adatoms are indicated. The hollow site, H: the adatom (grey/orange ball) is placed on top of the center of a hexagon. TN site: the adatom is placed on top of nitrogen (dark grey/red). TB site: the adatom is placed on top of boron (light grey/green). The bridge site BR: the adatom is located on top of the boron-nitrogen bond.}

\label{fig:structure} \end{figure} \par\end{center}

\section{Method}
We perform first-principles, spin-polarized plane wave calculations\cite{payne,vasp} within density functional theory (DFT)\cite{kohn} using PAW (projector augmented-wave) potentials.\cite{blochl} The exchange-correlation potential is approximated by generalized gradient approximation (GGA).\cite{gga} For the partial occupancies, we use the Methfessel-Paxton smearing method.\cite{methfessel} The width of the smearing for all structures is chosen as $0.01$ eV for geometry relaxations and band structure calculations. For accurate density of states (DOS) calculations the width of smearing is taken as $0.1$ eV. We consider a single adatom adsorbed to each (2x2) and (4x4) supercells of  2D BN structure and treat the system using periodic boundary conditions. For high coverage corresponding to $\Theta = 1/8$, (2x2) supercell is used , while low coverage, $\Theta = 1/32$, is treated by using (4x4) supercell. A large spacing (at least $\sim 14$ \AA) between adjacent 2D BN layers is taken to prevent interlayer interactions. The number of plane waves used in expanding Bloch functions and \textbf{k}-points used in sampling the Brillouin zone (BZ) are determined by a series of convergence tests. In the self-consistent potential and the total energy calculations, the BZ is sampled by (15x15x1) mesh points in \textbf{k}-space within Monkhorst-Pack scheme\cite{monk} for the (2x2) supercells. For calculations involving (4x4) supercells, the number of \textbf{k}-points is taken as (9x9x1). For accurate density of states (DOS) calculations, \textbf{k}-points samplings are further increased to (25x25x1) and (15x15x1) for the (2x2) and (4x4) supercells, respectively. A plane-wave basis set with the kinetic energy cutoff $\hbar^2 |\textbf{k}+\textbf{G}|^2/2m = 520$ eV is used.  All the atomic positions and lattice constants are optimized by using the conjugate gradient method, where the total energy and the atomic forces are minimized. The convergence is achieved when the difference of the total energies of last two consecutive steps is less than $10^{-5}$ eV and the maximum force allowed on each atom is 0.03 eV/\AA. The pressure on the system is kept smaller than $\sim 1$ kBar per unit cell in all of the calculations. As a rule, the structure becomes more energetic as its total energy is lowered.  Charge transfer values are calculated according to the Bader analysis.\cite{Bader}

\section{ADSORPTION}

The lowest energy sites of various adsorbed atoms are determined by placing foreign atoms initially to four possible adsorption sites at a height of $\sim 2$~\AA~ from BN plane as described in Fig. \ref{fig:structure}. Upon fully self-consistent geometry optimizations with both spin-polarized and spin-unpolarized configurations, where all atoms in the supercell are relaxed in all directions, we determine the equilibrium site as the lowest energy configuration among four different sites. The binding energy of an adsorbed adatom is defined as $E_{b} = E_{BN}+E_{A}-E_{BN+A}$, where $E_{BN}$ is the total energy of bare 2D BN, $E_{A}$ is the total energy of free adatom calculated in the vacuum. $E_{BN+A}$ is the total energy of 2D BN structure with adsorbed adatom.  We investigated the adsorption of following single adatoms, namely Sc, Ti, V, Cr, Mn, Fe, Mo, W, Pt, H, C, Si, B, N, O, Ca, Cu, Pd, Ni and Zn. Among these atoms, Cr, Mn, Mo, W, H, N, Ca and Zn cannot bind to 2D BN.

\subsection{Adsorption of Adatoms to (2x2) BN supercell ($\Theta = 1/8$)}

The electronic and magnetic properties of 2D BN are modified through adatom adsorption at high coverage. The adatom-adatom distance is $\sim$ 5 \AA~at $\Theta = 1/8$, where the magnetic coupling may be crucial in determining the ground state. In order to account for the antiferromagnetic (AFM) coupling between adatoms and to allow their reconstruction we treated $\Theta = 1/8$ coverage in (4x4) supercell, which includes four (2x2) supercells
each having a single adatom. This way adatom-adatom distance of $\sim$ 5 \AA~is maintained. We further carried out geometry relaxation with three different initial magnetic ordering of adatoms. First case is where adatoms are coupled antiferromagnetically. The second case is similar to first one, but adatoms are initially coupled ferromagnetically(FM). The last case corresponds to a spin-unpolarized, nonmagnetic (NM), geometry relaxation. Our calculations indicate that Cu, Ni, Pd and Pt adatoms have nonmagnetic ground state for $\Theta = 1/8$,  whereas Sc, Ti and V have AFM ground state. Oxygen is the only adatom which is found to be in the FM ground state. Boron and carbon adatoms undergo a reconstruction to lower the total energy in (4x4) supercell at $\Theta = 1/8$. Among all adatoms, Si and Fe present the most interesting situation, where 2D BN monolayer is changed to half-metallic state. However, these adatoms have small binding energies and thus are excluded from our further analysis. Also Cu having a binding energy smaller than 0.25 eV is also excluded from our study. Our results are given in Table \ref{tab:2x2}. In Figure ~\ref{fig:2x2-nonmagnetic}, the calculated energy band structures and corresponding total density of states (TDOS) and partial density of states (PDOS) of Ni, Pd or Pt adatom+2D BN system are presented. The band structure of bare 2D BN folded to (2x2) BZ is also presented to reveal the effect of the adatom adsorption on the electronic structure.

\begin{table*}
\caption{ Calculated structural, electronic and magnetic properties of 2D BN monolayer at uniform $\Theta = 1/8$ adatom coverage. Equilibrium positions of the adatoms, such as TN, TB, H and BR are described in Fig. \ref{fig:structure}; distances of adatom to the nearest N, $d_N$ (in \AA); distances of adatom to the nearest B, $d_B$ (in \AA); the height of adatom from the BN plane, $h$ (in \AA); the average B-N bond length, $d_{BN}$ (in \AA); the binding energy of adatom, $E_b$ (in $eV$); the net magnetic moment
per supercell, $\mu$ (in $\mu_B$). Electronic structure is specified as metallic (M) or semiconductor (SC). The type of the band gap can be either direct ($dr$) or indirect gap ($id$). The energy gap of the system after adsorption, $E_g$ (in $eV$); the transfer of charge to the adatom from BN, $\Delta \rho $ (in $electrons$ and $\Delta \rho <0$, if adatom is negatively charged); the dipole moment of the system along the $z-$ direction, p (in $e$\AA); work function (or photoelectric threshold for semiconductors) is $\Phi$ (in $eV$). For the adatoms, which have either nonmagnetic (NM) or ferromagnetic (FM) ground states, calculations are performed using (2x2) supercell; whereas for adatoms, which have either antiferromagnetic (AFM) ground state or reconstruction, (4x4) supercell are used with $\Theta = 1/8$. For adatoms undergoing a reconstruction, $\Delta \rho$ is given for the average charge transfer per adatom. }\label{tab:2x2} \centering
\begin{tabular}{cc|ccccccccccccc}
 Magnetic & Adatom & Position & $d_{N}$ & $d_{B}$ & $h$ & $d_{BN}$ & $E_{b}$ & $\mu$ & Electronic & Gap & $E_{g}$ & $\Delta \rho$ & p&  $\Phi$ \tabularnewline

 Ground State & &  & &  &  &  &  &  & Structure & Type &  &  &  & \tabularnewline
\hline \hline

\multirow{3}{*}{ NM }

 & Ni & TN & 1.87 & 2.31 & 1.84 & 1.46 & 1.14 & - & SC & $dr$ & 0.43 & 0.17 & -0.18 & 3.91\tabularnewline
\cline{3-3}
 & Pd & TN & 2.15 & 2.57 & 2.12 & 1.46 & 0.94 & - & SC & $id$ & 0.93 & 0.02 & -0.15 & 4.41\tabularnewline
 \cline{3-3}
 & Pt & TN & 2.04 & 2.49 & 2.04 & 1.46 & 1.48 & - & SC & $id$ & 0.93 & -0.03 & -0.10 & 4.66\tabularnewline
\hline \hline

\multirow{1}{*}{ FM }

& O & TB & 2.33 & 1.47 & 1.76 & 1.46 & 0.98 & 2.00 & SC & $id$ & 0.05 & -0.64 & 0.24 & 7.02\tabularnewline

\hline \hline

\multirow{3}{*}{ AFM }

& Sc & H & 2.49 & 2.54 & 2.03 & 1.46 & 0.90 & 0.00 & M & - & - & 0.67 & 0.30 & 3.03\tabularnewline \cline{3-3}
 & Ti & H & 2.44 & 2.48 & 1.99 & 1.45 & 0.92 & 0.00 & M & - & - & 0.52 & 0.98 & 3.56\tabularnewline
\cline{3-3}
 & V & TN & 2.21 & 2.67 & 2.26 & 1.45 & 0.60 & 0.00 & SC & $id$ & 0.19 & 0.36 & 0.04 & 2.81\tabularnewline

\hline \hline

Reconstruction

 & B & TN & 1.61 & 1.98 & 1.58 & 1.46 & 0.80 & - & SC & $id$ & 0.64 & 0.40 & 1.09 & 5.45\tabularnewline

\cline{3-3} NM & C & BR+TN & 1.52 & 1.66 & 1.28 & 1.46 & 1.38 & - & SC & $dr$ & 0.35 & -0.45 & 0.54 & 5.74\tabularnewline

\hline \end{tabular} \end{table*}

Nickel atom with $4s^{2}+3d^{8}$ electronic configuration is adsorbed at TN site. Even if Ni atom is placed at H or BR sites, it eventually moves to TN site. Flat bands slightly below the Fermi energy are all derived from the $3d$ orbitals of Ni atom. The band above the Fermi level is mainly formed from the combination of $4s$ and $3d_{z^{2}}$ orbitals of Ni. Accordingly, the band structure of parent BN layer is not distorted considerably. The charge transfer between Ni and 2D BN is relatively small and does not cause any
significant dipole moment.

\begin{center}
\begin{figure*}
\includegraphics[width=14 cm]{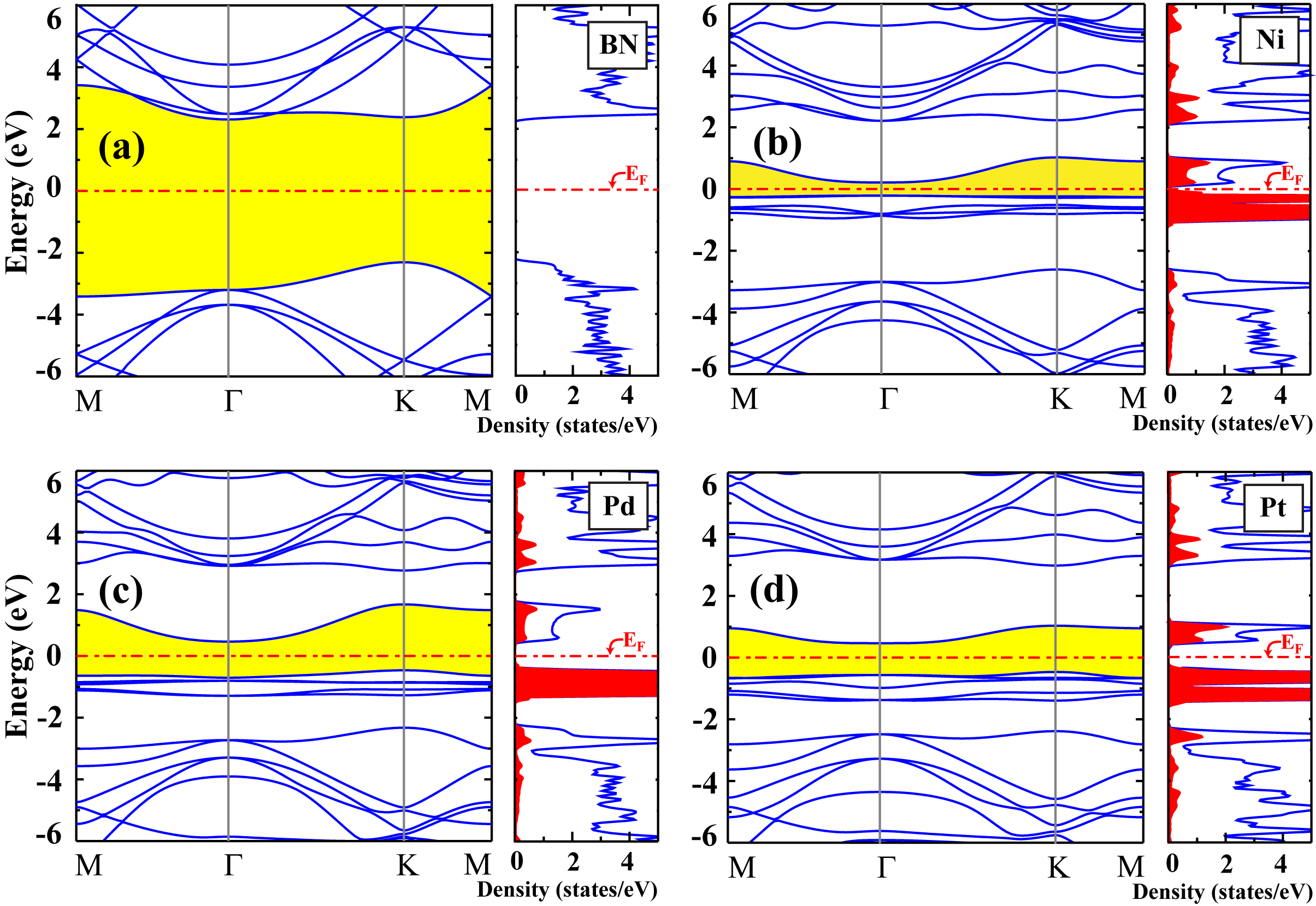}
\caption{(Color online) (a) The energy band structure of bare
2D BN folded to the (2x2) supercell and corresponding total density of states (TDOS). Zero of band energy is set at the Fermi energy,
$E_{F}$. (b)-(d) The energy band structure of single Ni, Pd, and Pt adsorbed to each (2x2) supercell of 2D BN ($\Theta = 1/8$).
Corresponding TDOS (continuous black/blue line) and partial density of states (PDOS) projected to the adatom (shaded dark/red) are
also indicated. The energy gap of semiconductors are shaded (light/yellow). The ground states are nonmagnetic.}
\label{fig:2x2-nonmagnetic} \end{figure*}

\par\end{center}

Adsorption of Pd and Pt give rise to electronic and magnetic properties similar to those of adsorbed Ni. They are adsorbed also at TN site and have high binding energies. Since the radius of Pd and Pt are relatively larger than Ni, the distance to the nearest N atom of 2D BN, $d_N$ is slightly larger. The charge transfer between 2D BN and the adatom is small. Ni, Pd and Pt have ionization energies of $7.63$, $8.33$ and $8.96$ eV, respectively.\cite{Kittel} This ordering of the ionization energies complies with the ordering of work function, $\Phi$, of Ni, Pd and Pt covered 2D BN. For both Pd and Pt, the bands slightly below the Fermi level are all derived from localized $d$ orbitals, but the band above Fermi level is mainly formed from $s$ orbital of adatom with some $d_{z^{2}}$ contribution similar with the case in Ni adsorption. Even if Pd and Pt have the same indirect band gap energy, Pd shows more dispersive band slightly above the Fermi level.

\begin{center}
\begin{figure}
\includegraphics[width=8cm]{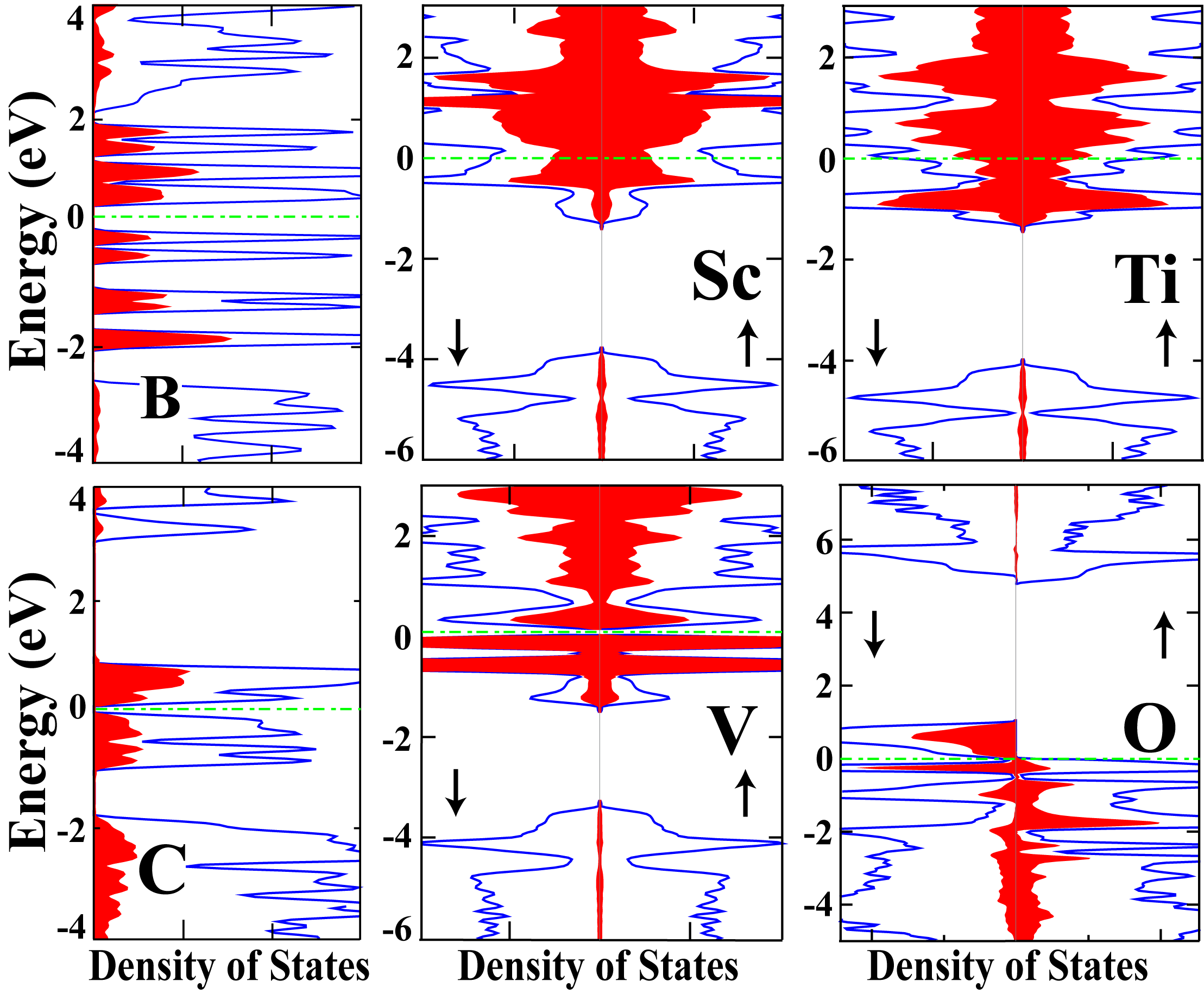}
\caption{(Color online) TDOS (dark/blue line) and PDOS (shaded grey/red) for B, C, O, Sc, Ti and V adatoms at high coverage, $\Theta =1/8$. Up (down) arrow on the right (left) site of TDOS and PDOS indicates spin direction. Zero of the DOS energy is set to the Fermi
energy, $E_{F}$ of the adatom+2D BN system which is indicated with dashed-dotted grey/green line. All adatoms treated in (4x4)
supercell to allow antiferromagnetic interaction or reconstruction, whereby single adatom is adsorbed to each (2x2) supercells of
(4x4) supercell amounting $\Theta = 1/8$. Sc, Ti and V have antiferromagnetic ground state; O has ferromagnetic state; B and C are
nonmagnetic. The state densities are given in arbitrary units, but have the same scale for all adatoms.}

\label{fig:2x2-magnetic} \end{figure}

\par\end{center}

TDOS and PDOS corresponding to $\Theta = 1/8$ coverage of B, C (which undergo a reconstruction in (4x4) supercell); Sc, Ti, V (which have AFM ground state), and O (which has FM ground state) are presented in Fig.~\ref{fig:2x2-magnetic}. Four carbon adatoms treated in a (4x4) supercell undergo a reconstruction\cite{CBreconstruction}; while three C atoms are adsorbed near TN sites, the remaining one
is moved to the BR site. Upon reconstruction, the charge on C adatom at the bridge cite is minute positive, whereas those at TN site are negatively charged by $\sim 0.6$. Carbon adatoms at TN site take their charge mostly from the nearest N atom. The bonding of C at TN cite has an ionic character resulting a dipole moment on the system.

Sc, Ti and V have AFM ground state at $\Theta = 1/8$.\cite{AFM} Unlike Sc and Ti, adsorption of V does not make the system metallic. The energy bands below the Fermi level is mainly from $d_{xy}, d_{yz},d_{xz}$ and above the Fermi level is mainly from $d_{x^2}$ and $d_{z^2}$. The energy band gap of 0.19 eV originates from the splitting of $3d$ orbital states.

Among all adatoms in Table \ref{tab:2x2}, only O has FM ground state and is absorbed at TB site. The excess charge on the adsorbed O is transferred from three nearest N atoms, which leads to an ionic character in bonding. Since underlying B atom is positively charged and O is negatively charged, adsorbed O pulls the nearest B atom and causes distortion on planar 2D BN monolayer.

\subsection{Adsorption of Adatoms to (4x4) BN supercell ($\Theta = 1/32$)}

We next investigate the adsorption of a single adatom to (4x4) supercell, which corresponds to low density coverage, $\Theta = 1/32$. As indicated in Fig. \ref{fig:structure}, we initially placed the adatoms at four different positions and relaxed their geometric structure with both spin-polarized and spin-unpolarized calculations. Interestingly, we observe that among 9 different adatoms, the adsorption sites of B, C and O differ by going from $\Theta = 1/8$ to $\Theta = 1/32$. Since the adatom-adatom distance between adjacent supercells is $\sim 10$~\AA, it can normally be contemplated that the coupling between adatoms are negligible. To verify this assumption, we carried out binding energy calculations for C, O, Ti, Sc and Pt adatoms in (8x8) supercell corresponding to a coverage of $\Theta = 1/128$. In Table \ref{tab:8x8} we compared the equilibrium binding sites and binding energies, and magnetic states of these atoms for three different coverage $\Theta = $1/8, 1/32 and 1/128.

\begin{table}
\caption{Binding energies ($E_b$), magnetic moments ($\mu$), adsorption sites and average adatom-adatom distances\cite{average}
($\overline{d}_{A-A}$) of C, O, Sc, Ti and Pt adatoms adsorbed on 2D BN at different coverages.} \label{tab:8x8}
\begin{tabular}{cc|ccccc} $\Theta$ & & C & O &Sc  & Ti  & Pt \tabularnewline \hline

\multirow{2}{*}{ 1/8 } & $E_{b}$ (eV) & 1.38 & 0.98 & 0.90 & 0.92 & 1.48 \tabularnewline & $\mu$ ($\mu_{B}$) & - & 2.00 & 0.00 & 0.00
& -\tabularnewline $\overline{d}_{A-A} = \sim5$~\AA & Site & BR+TN & TB & H& H & TN \tabularnewline \hline

\multirow{2}{*}{ 1/32 } & $E_{b}$ (eV) & 1.17 & 2.01 & 0.43 & 0.74 & 1.52 \tabularnewline & $\mu$ ($\mu_{B}$)& 2.00 & - & 3.00 & 4.00
& - \tabularnewline $\overline{d}_{A-A} = \sim10$~\AA & Site & BR & BR & H & H & TN \tabularnewline

\hline

\multirow{2}{*}{ 1/128 } & $E_{b}$ (eV) & 1.15 & 1.98 & 0.37 & 0.70 & 1.47 \tabularnewline & $\mu$ ($\mu_{B}$) & 2.00 & - & 3.00&
4.00 & - \tabularnewline $\overline{d}_{A-A} = \sim20$~\AA & Site & BR & BR & H & H & TN\tabularnewline

\hline \hline

\end{tabular} \end{table}

The binding energies as well as the magnetic ground states of C, O, Sc and Ti vary substantially by going from $\Theta = 1/8$ to $\Theta = 1/32$. For C and O, the adsorption sites are also changed. However the situation is not the same when the coverage is further lowered from $\Theta = 1/32$ to $\Theta = 1/128$, hence when the adatom-adatom distance is increased from $\sim$ 10 \AA~ to $\sim$ 20 \AA. For adatoms included in Tab. \ref{tab:8x8}, the binding energies and magnetic moments are not changing significantly, and the adsorption sites are remaining the same by going from $\Theta = 1/32$ to $\Theta = 1/128$. This finding is corroborating our arguments made at the beginning of the paper, that the properties calculated for $\Theta = 1/32$ coverage (or the adsorption of a single adatom adsorbed to each (4x4) supercell) can mimic the adsorption of single isolated adatom on a very large area of 2D BN (or very large adatom-adatom distance). Consequently, the bands of adatom+2D BN calculated at $\Theta =1/32$ become rather flat and can be taken as the localized impurity state (or resonances if 2D BN states are significantly contributed). Under these circumstances, the band gap and the edges of valence and conduction bands can be unaltered. In Table \ref{tab:4x4}, we include the calculated geometric, electronic and magnetic properties of a single adatom adsorbed to 2D BN at $\Theta = 1/32$. The energies of localized and resonance states relative to 2D BN's valence band edge are also tabulated.

\begin{widetext}
\centering
\begin{table*}
\caption{Calculated values for single adatom adsorbed to each (4x4) supercell, corresponding to the coverage $\Theta = 1/32$. Adsorption site, binding energy, $E_b$; magnetic moment per (4x4) supercell, $\mu$; the distance from 2D BN monolayer, $h$; the distance from the nearest N atom, $d_N$; the distance from the nearest B atom, $d_B$; charge transfer, $\Delta \rho$; and energies of relevant localized or resonance states measured from the top of the valence band, E$_n$. (F$\uparrow$) indicates that the corresponding spin-up state is full. (E$\downarrow$) represents an unoccupied spin-down state. If no spin direction is indicated, that state is nonmagnetic. R is resonance state having significant contribution from 2D BN states in the band continua. Adsorption site of the adatoms corresponding to their lowest total energy are indicated by TN, TB, BR, or H as described in Fig.\ref{fig:structure}.}

\label{tab:4x4}. \begin{tabular}{c|cccccccccccc}

&Ni & Pd & Pt & C & Sc & Ti & V & B & O  \tabularnewline \hline

Site & TN & TN &TN & BR & H & H & TN & BR & BR  \tabularnewline

E$_{b}$ (eV) & 1.15 & 0.96 &  1.52 & 1.17 & 0.43 & 0.74 & 0.58 & 0.79 & 2.01  \tabularnewline

$\mu$ ($\mu_{B}$) & - & - & - & 2.00 & 3.00 & 4.00 & 5.00 & 1.00 & -  \tabularnewline

$h$ (\AA) & 1.87 & 2.15 & 2.14 & 1.71 & 2.08 & 2.04 & 2.28 & 1.83 & 1.56  \tabularnewline

$d_{N}$ (\AA) & 1.86 & 2.14 & 2.02 & 1.59 & 2.51 & 2.46 & 2.21 & 1.63 & 1.52 \tabularnewline

$d_{B}$ (\AA) & 2.31 & 2.56 & 2.50 & 1.77 & 2.53 & 2.49 & 2.66 & 1.79 & 1.48 \tabularnewline

$\Delta\rho $ & 0.16 & 0.01 & -0.01 & -0.23 & 0.59 & 0.42 & 0.29 & 0.30 & -0.81 \tabularnewline

\hline \hline

E$_1$ &1.43 (F)& 0.65 (F) & 0.32 (F) & 1.11 (F $\uparrow$) & 2.72 (F $\uparrow$) & 2.73 (F $\uparrow$) & 2.16 (F $\uparrow$) & 0.26 (F $\uparrow$) & -0.21 (FR)  \tabularnewline

E$_2$ &1.83 (F)& 0.92 (F) & 0.61 (F) & 1.61 (F $\uparrow$) & 3.14 (F $\uparrow$) & 3.35 (F $\uparrow$) & 2.57 (F $\uparrow$) & 0.70 (F
$\downarrow$) & -0.12 (FR)  \tabularnewline

E$_3$ &2.80 (E)& 1.18 (F) & 0.79 (F) & 2.39 (E $\downarrow$) & 3.75 (E $\downarrow$) & 3.97 (E $\uparrow$) & 3.01 (F $\uparrow$) &
2.02 (F $\uparrow$) & 0.49 (FR)  \tabularnewline

E$_4$ &4.49 (ER)& 2.95 (E) & 2.25 (E) & 3.38 (E $\downarrow$) & 3.77 (E $\uparrow$) & 4.06 (E $\downarrow$) & 3.48 (E $\uparrow$) &
3.02 (E $\downarrow$) & 0.66 (FR)\tabularnewline

E$_5$ & & & & & 3.96 (E $\downarrow$) & 4.74 (E $\downarrow$) & 3.88 (E $\downarrow$) & 3.18 (E $\downarrow$) & 4.14 (ER) &
\tabularnewline

E$_6$ & & & & & 4.01 (E $\uparrow$) & & 4.32 (E $\downarrow$) & 3.42 (E $\uparrow$) & & \tabularnewline

E$_7$ & & & & & 4.49 (E $\downarrow$) & & 4.77 (E $\downarrow$) & & \tabularnewline


\hline
\end{tabular}
\end{table*}
\end{widetext}

Boron adsorbed on 2D BN on (4$\times$4) supercell exhibits electronic, magnetic and structural properties, which are significantly different from those of $\Theta = 1/8$. For example, B atom of the planar BN rises $\sim 0.2$ \AA; this slightly changes the $sp^{2}$-hybridization locally. Two spin-up bands and one spin-down band for spin-down are filled below the Fermi energy. In Figure \ref{fig:4x4-band}, the band structure of B+2D BN and isosurfaces of charge density corresponding to the localized states E$_1$, E$_3$ and E$_5$ are shown. The magnetic properties are different from the case at $\Theta = 1/8$, since reconstruction of B adatoms does not take place due to the absence of adatom-adatom interaction.

Adsorption of C on (4$\times$4) supercell of 2D BN causes significant splitting of degenerate $p$ orbital levels. Spin-up and spin-down bands indicated in Fig. \ref{fig:4x4-band} originate from $p_{x}$ (E$_1$ state) and $p_{y}$ (E$_2$ state) orbitals of adsorbed C atom. In this low coverage, C adatom is adsorbed to BR position, but closer to N atom. Similar to B, the magnetic properties of the C+BN system changed from NM to FM upon lowering the coverage density.

\begin{center}
\begin{figure*}
\includegraphics[width=16cm]{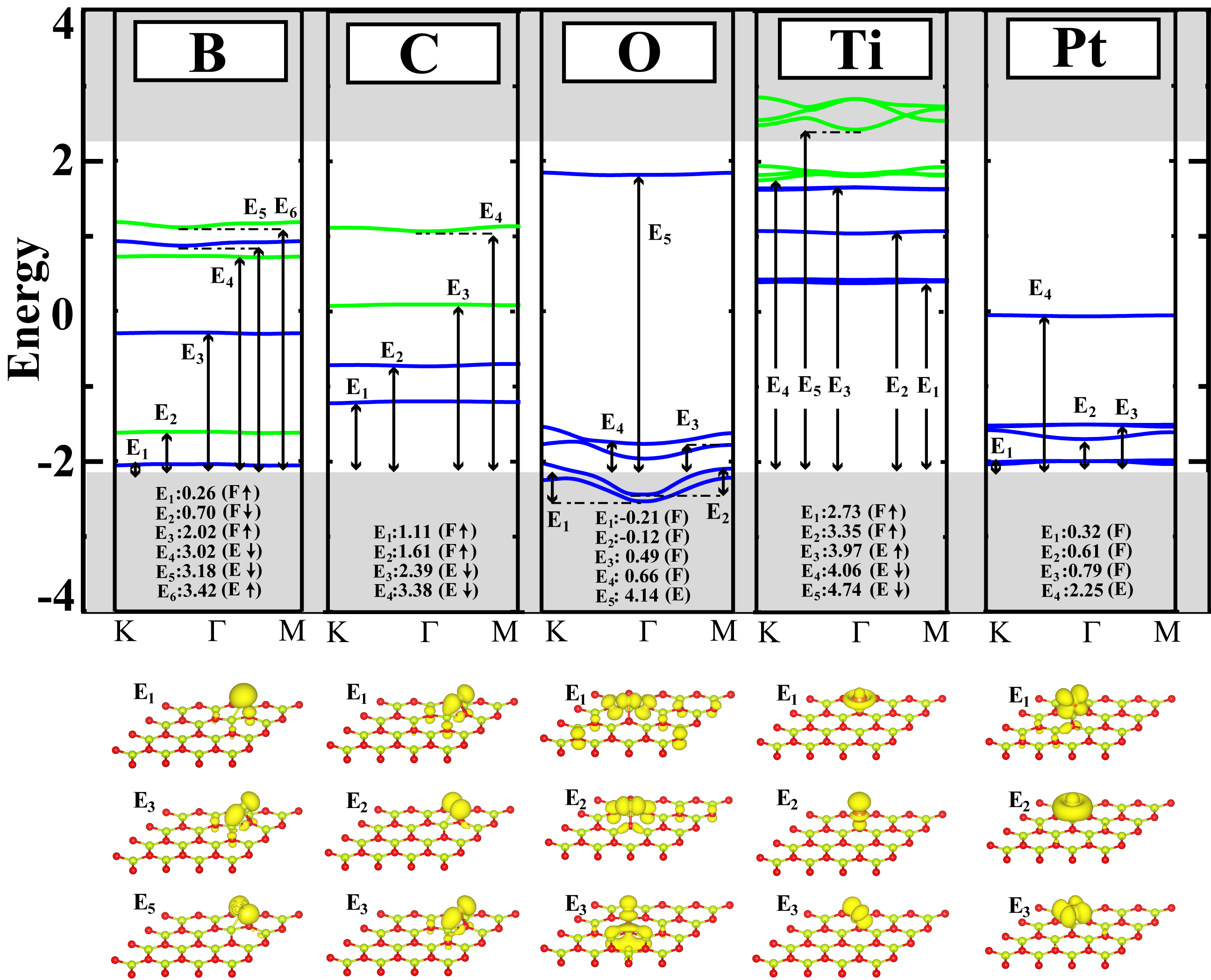}
\caption{(Color online) Schematic diagram of the relevant energy levels (or bands) of adatom (B, C, O, Ti and Pt) adsorbed to
(4$\times$4) supercell of the single layer BN. The light/grey shaded region in the background is the valence and conduction band
region of 2D BN. Zero of band energy is set at the Fermi level of the parent, bare 2D BN, $E_{F}$. Spin-up and spin-down bands are
shown by dark/blue and light/green lines, respectively. Solid bands indicates that the contribution of adatom to the band is more than 50\% except O adatom. Energies of some of the relevant adatom (impurity) states relative to valence band edge of parent 2D BN are indicated. Below each energy diagram of adatoms, charge density isosurfaces of specific states are shown. The isosurface value is taken as 7x10$^{-5}$ electrons/\AA$^3$. (Note that all localized states of O have significant contributions from BN states and are shown by solid dark/blue lines.)}
\label{fig:4x4-band} \end{figure*}
\par\end{center}

At $\Theta = 1/8$, O adatom creates a surface distortion and pulls underlying B atom upwards. However at $\Theta = 1/32$, O adsorbed at BR site does not generate a distorted BN region. The bands near the edge of valence band have significant dispersion due to strong coupling between O and 2D BN. The AFM ground states of both Sc, Ti and V adsorbed 2D BN at $\Theta = 1/8$ change to FM ground states at $\Theta = 1/32$. Among all adatoms studied in the present paper, electronic and magnetic properties of Ni, Pd and Pt are not affected upon lowering the coverage density.

\begin{figure}
\centering
\includegraphics[width=6cm]{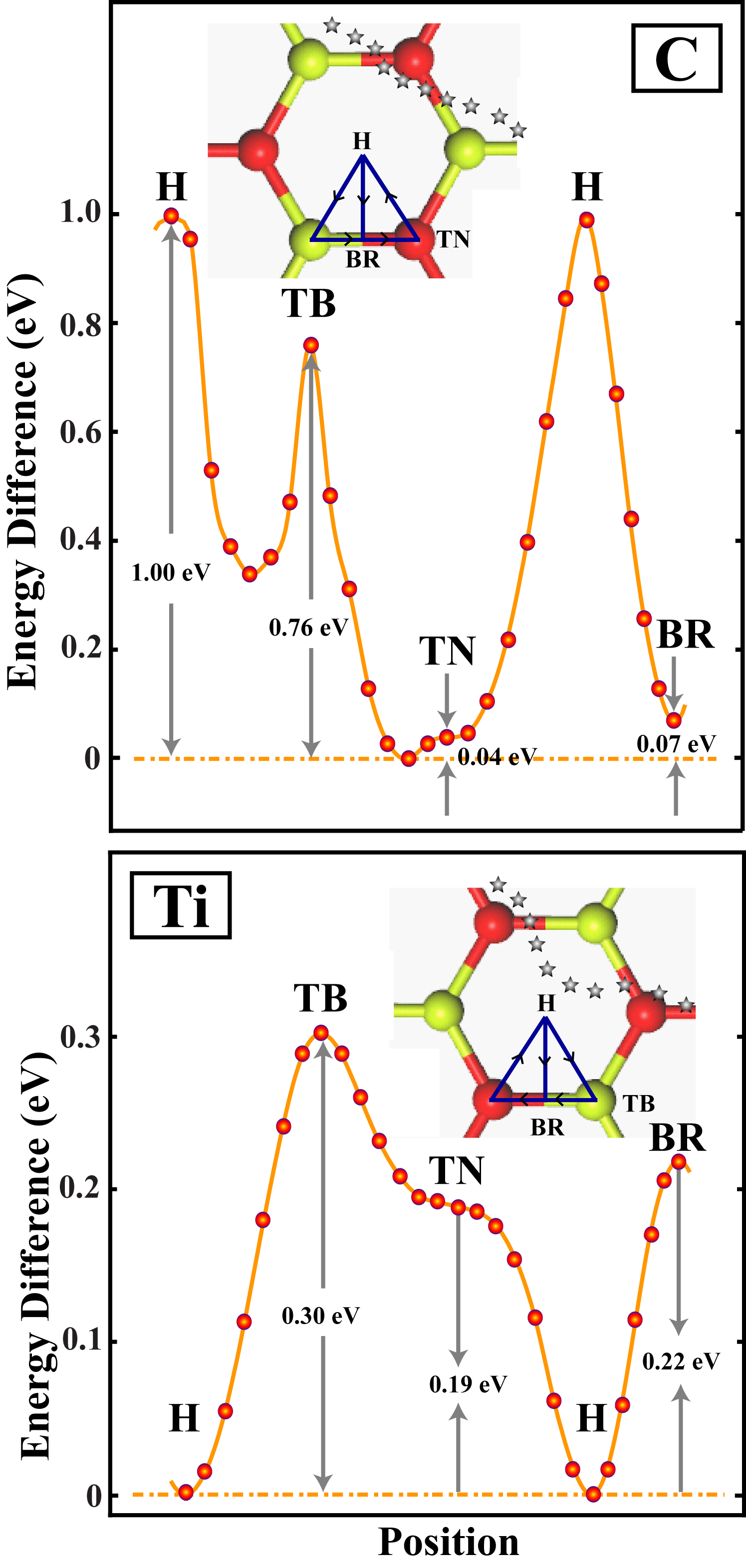}
\caption{(Color Online) Variation of energy of a single C and Ti
adatom as they are moving along special directions of a hexagon of 2D BN honeycomb structure. The possible migration path of adatom
facing minimum energy barrier is shown by stars on the hexagon. Calculations are carried out in (4x4) supercell.}
\label{fig:diffusion} \end{figure}

Next we address the question whether adsorbed adatom migrates on 2D BN honeycomb structure. Here we consider only single C and Ti atoms as prototypes and calculate the energy barrier for their motion along the symmetry directions of 2D BN. In Figure \ref{fig:diffusion}, the variation of energy of single C and Ti adatoms along symmetry directions are presented. The potential barrier on the migration path for C adatom is $\sim$0.30 eV. In the case of Ti the energy barrier on the possible migration path is only 0.15 eV. The latter barrier is not high enough to prevent Ti atoms from cluster formation at elevated temperatures.

\section{Substitution ($\Theta = 1/32$)}

We, finally deal with the substitutional doping of the specific atoms. Here we consider that Al, Be and C substitution for B atom; C, O, and P substitution for N atom of 2D BN honeycomb structure. We mimic the substitutional doping by a model, where a specific atom substitutes for a single B or N in every (4x4) supercell of 2D BN. In this model the distance between impurity atoms is $\sim$ 10~\AA~ resulting in negligible coupling between them. The substitution energy $E_{s}$ is calculated\cite{formula1, formula2} as:

\begin{equation*} \centering E_{s} ~~~=~~~\frac{N-1}{N} E_{BN} + E_{A} - E_{B\bigvee N} - E_{BNS} \end{equation*}

Here $N$ is the number of atoms in the supercell, which is $N=32$ for (4x4) supercell. $E_{BN}$ is the calculated total energy of 2D BN of the (4x4) supercell corresponding to 16 B-N atom pairs. $E_{A}$ is the experimental cohesive energy\cite{Kittel} of the impurity atom in its equilibrium crystal. $E_{BNS}$ is the total energy of the supercell after substitution process. $E_{B\bigvee N}$ is the cohesive energy of either B or N that depends on which atom is exchanged. The experimental cohesive energy of B is defined with respect to the B crystal.\cite{Kittel} However, in the case of N, the cohesive energy of N is calculated relative to N$_2$ in the gas phase. According to this expression, a positive energy value means the exchange of foreign adatom either with B or N is endothermic reaction, whereas the negative value indicates an exothermic process. Structural relaxations and the lowest energy states of the final structures are further tested with three different initial magnetic moment distributions on the atoms and all these cases are converged to the same values given in Tab. \ref{tab:subs}. Here the substitution of C atom for B (N) is of particular interest, since it dopes the 2D BN honeycomb structure as donor (acceptor).

\begin{widetext}
\centering
\begin{table*}
\caption{Substitution energy, $E_s$; magnetic moment per (4x4) supercell, $\mu$; the height of the substituting atom from 2D BN monolayer, $h$; the distance from the nearest B or N  atom, $d_{B \bigvee N}$; charge transferred to substituting atom\cite{Bader}, $\Delta \rho$; and energies of relevant localized states in the band gap, $E_n$. (F$\uparrow$) indicates that the corresponding spin-up state is full. (E$\downarrow$) indicates unoccupied spin-down state. If no spin direction is indicated, that
state is nonmagnetic.} \label{tab:subs}

\begin{tabular}{c|ccc||ccc} \multicolumn{1}{c}{} & \multicolumn{3}{c||}{B$\rightarrow$} & \multicolumn{3}{c}{N$\rightarrow$}
\tabularnewline \cline{2-7}

& Al & Be & C & C & O & P \tabularnewline \hline

E$_{s}$ (eV) & 5.47 & 5.23 & 5.47 & 3.42 & 6.51 & 3.77 \tabularnewline

$\mu$ ($\mu_B$) & NM & NM & 1.00 & 1.00 & 1.00 & NM \tabularnewline

$h$ (\AA) & 0.52 & 0.01 & 0.01 & 0.00 & 0.00 & 1.42 \tabularnewline

$d_{B \vee N}$ & 1.71 & 1.56 & 1.41 & 1.52 & 1.50 & 1.88 \tabularnewline

$\Delta\rho$ & 2.28 &1.60& 1.15 &-1.89 & -1.51 & -0.47 \tabularnewline


\hline \hline
E$_{1}$ & 2.76 (E) & -2.29 (F) & 3.50 (F $\uparrow$)   & -1.69 (F $\uparrow\downarrow$)    & -2.45 (F $\uparrow$)    &
-0.76 (F)   \tabularnewline
E$_{2}$ & 3.02 (E) & -0.83 (F) & 4.22 (E $\downarrow$) & -0.69 (F $\uparrow\downarrow$)    & -2.16 (F $\downarrow$)  &  -0.07 (F)   \tabularnewline
E$_{3}$ & 4.22 (E) &  4.38 (E) &   &  0.17 (F $\uparrow$)   &    &  4.20 (E)   \tabularnewline
E$_{4}$ & 6.26 (E) & & & 1.07 (E $\uparrow$)  &                         &                        \tabularnewline

\end{tabular}

\end{table*} \end{widetext}

\begin{center}
\begin{figure*}
\includegraphics[width=16cm]{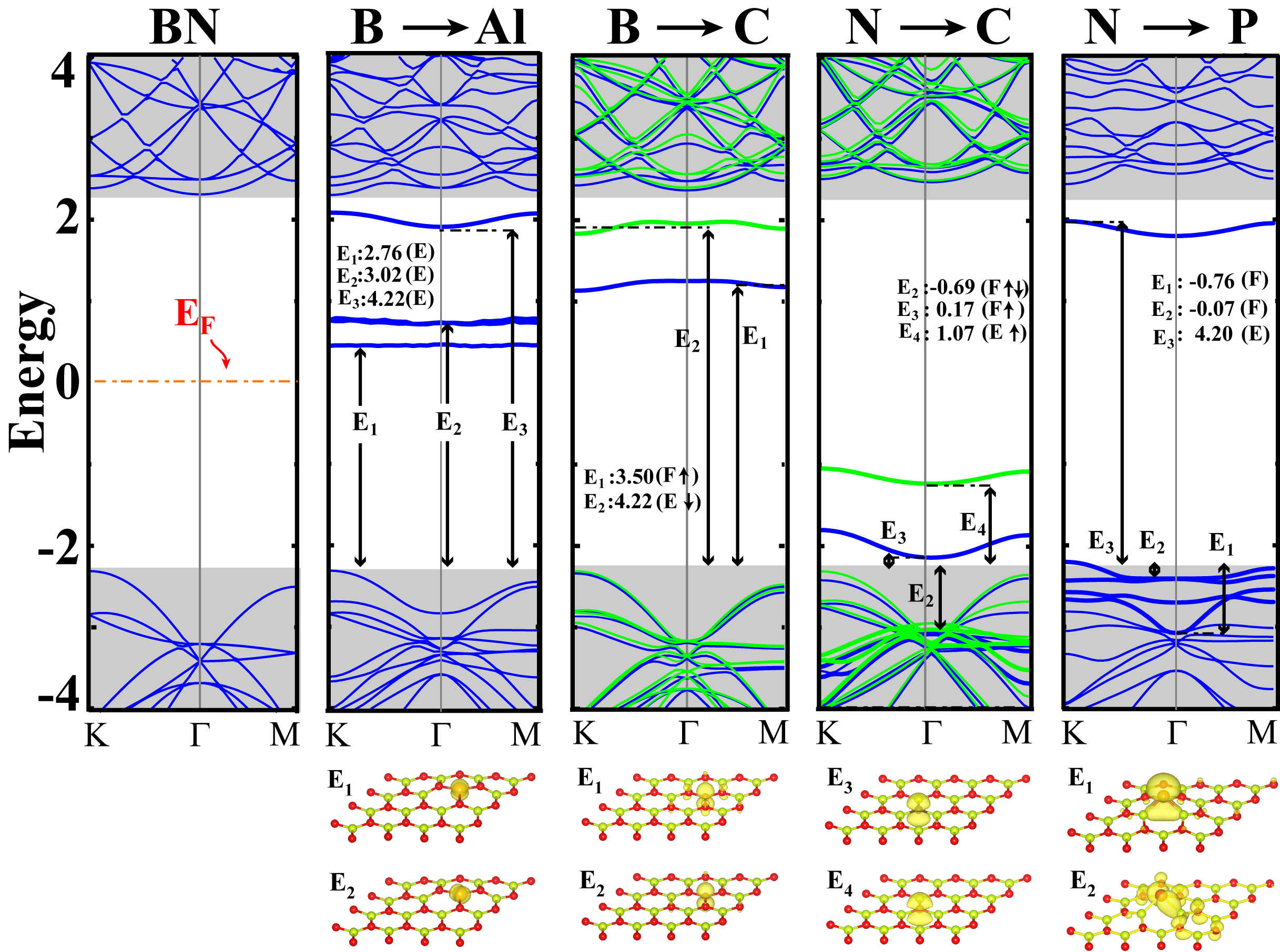}
\caption{(Color online) Schematic energy level diagram of Al and C substituting B; and C and P substituting N in a (4x4) supercell of
2D BN. The light/grey shaded region in the background is the valence and conduction band continuum of 2D BN. Zero of energy is set to
the Fermi level of bare 2D BN. Spin-up bands are indicated with dark/blue lines, whereas spin-down bands are light/green. Thick solid bands (levels) are states localized at the substituted atom and nearest atoms of 2D BN. Thin solid bands stand for the delocalized states. The energies of some of the localized states with respect to the valence band edge are shown in units of eV. Charge density isosurfaces of some of the relevant states are shown below the corresponding band diagram. The isosurface value is taken as 7x10$^{-5}$ electrons/\AA$^3$.} \label{fig:substitution}
\end{figure*}
 \par\end{center}

We now examine the electronic and magnetic structure of 2D BN substituted by Al, C, and P described schematically in  Fig. \ref{fig:substitution}. Al substituting for B and being in the same group with B, but having a relatively large atomic radius, distorts the planar structure of BN layer and is located at a position $0.52$~ \AA~ higher than the substituted B atom. The valence band of the BN layer does not influence much upon substitution, but states derived from $p$-orbitals of Al adatom appear near the conduction band edge. In Figure \ref{fig:substitution}, $p_z$-orbitals of Al are dominant in E$_1$ state, however $p_y$ is the most contributed in E$_2$ state. In the case of C substituting B, the valence and conduction bands of BN layer are not influenced, except minute splitting between the spin-up and spin-down states resulting from the magnetic C atom. The excess electron of the substituted C atom relative to B leads spin-polarization and fills the E$_1$ state in Fig. \ref{fig:substitution} like a $n$-type semiconductor. When substitute for N, C atom creates electron deficiency; states derived from $p$-orbitals of C atom are located near the edge of the valence band like a $p$-type semiconductor. $p_z$ orbital of C splits and spin-up (E$_3$ state in Fig. \ref{fig:substitution}) becomes occupied. Because of this unpaired electron C atom attains net $1 \mu_{B}$ total magnetic moment in both $n$-type and $p$-type substitutional doping.

Substituted P having relatively larger atomic radius, gives rise to the local deformation and raises $1.42$ \AA ~ above the plane of 2D BN honeycomb structure. This gives rise to a local dehybridization of planar $sp^2$-bonding. Since the upper valence bands of parent 2D BN are derived mainly from nitrogen atoms, these bands are affected upon their exchange with foreign atoms. Hence several resonances appear, such as E$_1$ and E$_2$ in the upper part of valence band when P substitutes for N atom in  Fig. \ref{fig:substitution}. In addition a localized state E$_3$ occurs near the conduction band edge.

\section{Discussion and Conclusions}

In this paper, we showed that 2D BN can be functionalized to attain properties, which can be useful in future applications in nanoelectronics and nanomagnetics. Functionalization can be achieved either through the adsorption of foreign atoms at different coverage or substitution of foreign atoms for B or N in honeycomb structure. We considered a number of foreign atoms, such as Sc, Ti, V, Cr, Mn, Fe, Mo, W, Pt, H, C, Si, B, N, O, Ca, Cu, Pd, Ni and Zn. Part of these atoms are bound with a significant energy and form chemical bonds with 2D BN. Cu, Fe and Si have binding energy smaller than 0.25 eV, but Cr, Mn, Mo, W, H, N, Ca and Zn cannot bind to 2D BN. Owing to the van der Waals interaction, the true binding energies can be $0.1-0.2$ eV larger than those calculated with GGA in the present work.

High coverage of adatoms corresponding to $\Theta = 1/8$ leads to dramatic modifications in electronic structure, if the related binding energy is significant. Under these circumstances, either the wide band gap of 2D BN can be reduced or diminished and the system becomes metallic. Under certain circumstances, the nonmagnetic 2D BN attains magnetic moment. Ni, Pd, and Pt covered 2D BN are nonmagnetic semiconductors with band gap relatively smaller than that of the parent 2D BN. Sc, Ti forming a (2x2) structure on 2D BN are AFM metals, V+2D BN is an AFM semiconductor. Remarkably, B+ and C+2D BN undergo a (4x4) reconstruction and have a band gap smaller than the parent 2D BN. Oxygen covered 2D BN is found to be a FM, small band gap semiconductor, which may display high spin-polarization under bias voltage. If the interaction between adatom and 2D BN is weak, the electronic structure can be viewed as the combination of electronic band structure of 2D BN monolayer and adatom monolayer. Si and Fe are weakly bound to 2D BN and change the wide band gap of 2D BN to a half-metal.

 At low coverage corresponding to $\Theta = 1/32$, the large distance between adatoms hinders any significant interaction between them. This situation is taken to mimic a single, isolated atom adsorption to 2D BN, which gives rise to localized states in the wide band gap of 2D BN honeycomb structure. This conjecture is confirmed by examining the adsorption of single C, O, Sc, Ti and Pt to (8x8) supercell resulting in $\sim$ 20 \AA~adatom-adatom distance. In this respect, the adsorbed adatoms at $\Theta = 1/32$  act as dopants of 2D BN. In a few cases the adsorption site and magnetic state undergo a change by going from the high, $\Theta = 1/8$ to low coverage, $\Theta = 1/32$.

Not only adatoms, but also substitution of foreign atoms for B or N in the honeycomb structure give rise to localized impurity states in the gap, which attribute interesting electronic properties to the system. In particular, the substitution of C for B yields an excess charge and gives rise to two donor states near the edge of conduction band, the lower lying spin-up band being full. In contrast, the substitution of C for N yields a single-electron deficiency and gives rise to two acceptor states above the top of valence band, the lower lying spin-up state being full. In both cases, the unpaired spins of C atom give rise to magnetic moment $\mu = 1 \mu_B$.

In conclusion, while 2D BN is a mechanically stiff and nonmagnetic wide band gap semiconductor, its band gap can be engineered through adatom decoration. In specific cases 2D BN attains magnetic properties and becomes metallic. Foreign atoms adsorbing at low coverage or exchanging with B or N atoms give rise to donor like or acceptor like states in band gap. At the end, the material achieves interesting properties. Some of these properties can be exploited in future applications.

\section{Acknowledgement}

Part of computational resources have been provided through a grant (2-024-2007) by the National Center for
High Performance Computing, Istanbul Technical University. We thank the DEISA Consortium (www.deisa.eu), funded through the EU FP7 project RI-222919, for support within the DEISA Extreme Computing Initiative. SC acknowledges partial support from TUBA, Academy of Science of Turkey. Authors thank Dr. Ethem Akt\"{u}rk for helpful discussions.


\begin{thebibliography}{1}
\bibitem{CBreconstruction}
When treated in the (2x2) supercell at $\Theta = 1/8$, C adatoms attain net magnetic moment of 2 $\mu_B$ per (2x2) supercell. This configuration is, however, 0.3 eV per adatom higher (energetically less favorable) than those reconstructed in the (4x4) supercell. Similarly, for $\Theta = 1/8$ B treated in (2x2) supercell, binding at the BR site, attain magnetic moment of 1 $\mu_B$, but becomes 0.12 eV per adatom less favorable relative to the reconstructed structure.
\bibitem{AFM}
The excited state of AFM ground state of Sc, Ti and V is the FM state with the magnetic moment of 0.45 $\mu_{B}$, 3.51 $\mu_{B}$, and 4.80 $\mu_{B}$; and with the energies of 30 meV, 50 meV and 20 meV above the ground state, respectively
\bibitem{average} The optimized lattice parameters can be different for different adatoms and for different coverages.
\end{thebibliography}
\end{document}